\documentclass[12pt]{article}
\usepackage{epsfig}
\usepackage{amssymb}
\usepackage{amsmath}
\usepackage{amsfonts}

\newcommand{\R}{\mathbb{R}}
\newcommand{\C}{\mathbb{C}}

\newcommand{\fU}{\mathfrak{U}}

\newcommand{\be}{\begin{equation}}
\newcommand{\ee}{\end{equation}}
\newcommand{\bea}{\begin{eqnarray}}
\newcommand{\eea}{\end{eqnarray}}

\newcommand{\ed}{\end{document}}

\newcommand{\np}{\newpage}

\begin{document}

\title{Is Weak Pseudo-Hermiticity Weaker than Pseudo-Hermiticity?}
\author{\\
Ali Mostafazadeh
\\
\\
Department of Mathematics, Ko\c{c} University,\\
34450 Sariyer, Istanbul, Turkey\\ amostafazadeh@ku.edu.tr}
\date{ }
\maketitle

\begin{abstract}
For a weakly pseudo-Hermitian linear operator, we give a spectral
condition that ensures its pseudo-Hermiticity. This condition is
always satisfied whenever the operator acts in a
finite-dimensional Hilbert space. Hence weak pseudo-Hermiticity
and pseudo-Hermiticity are equivalent in finite-dimensions. This
equivalence extends to a much larger class of operators. Quantum
systems whose Hamiltonian is selected from among these operators
correspond to pseudo-Hermitian quantum systems possessing certain
symmetries. \vspace{5mm}

\vspace{2mm}


\end{abstract}


\np

\section{Introduction}

Refs.~\cite{p1,p23} discuss a notion of a pseudo-Hermitian
operator that has proven to be a convenient tool in the study of
${\cal PT}$-symmetric Hamiltonians \cite{pt} --
\cite{znojil-pla-2006}. It also plays a central role in solving
some of the basic problems of relativistic quantum mechanics and
quantum cosmology \cite{qc} and revealing some interesting
analogies between quantum mechanics and general relativity
\cite{pla-2004}. The following is a mathematically precise
description of this notion \cite{p1}.
\begin{itemize}
\item[]{\bf Definition~1:} A densely-defined linear operator
$H:{\cal H}\to{\cal H}$ acting in a separable Hilbert space ${\cal
H}$ is said to be \emph{pseudo-Hermitian} if there exists a
Hermitian automorphism $\eta:{\cal H}\to{\cal H}$ satisfying
    \be
    H^\dagger=\eta\,H\eta^{-1},
    \label{ph}
    \ee
where $H^\dagger$ denotes the adjoint of $H$.\footnote{Throughout
this paper, ``Hermitian'' means ``self-adjoint,'' e.g.,
$\eta^\dagger=\eta$.}
\end{itemize}
For a discussion of the earlier uses of the term pseudo-Hermitian
in the context of indefinite-metric theories see \cite{cjp-2003}.

Note that an automorphism is by definition an everywhere-defined,
one-to-one, and onto linear operator. Moreover, an
everywhere-defined Hermitian linear operator is necessarily
bounded\footnote{This is known as the Hellinger-Toeplitz theorem
\cite{reed-simon}.}, and a bounded one-to-one onto linear map has
a bounded inverse.\footnote{This is known as the inverse mapping
theorem \cite{reed-simon} or Banach's theorem
\cite{kolmogorov-fomin}.} As a result, if one adopts the
definition of an invertible operator that identifies the latter
with a one-to-one, onto linear map with a bounded inverse
\cite{hislop-sigal,schechter,halmos}, then \emph{a linear operator
is everywhere-defined, Hermitian, and invertible if and only if it
is a Hermitian automorphism}. Usually in physics literature one
ignores the technical issues associated with the domain of the
operators and uses ``Hermitian automorphism'' and ``Hermitian
invertible linear map'' synonymously. Another more familiar term
used for such an operator particularly in the context of
pseudo-Hermitian operators is ``\emph{pseudo-metric}''.

The operator equation~(\ref{ph}) in particular implies that the
domain of its both sides must coincide. In light of the fact that
$\eta$ is everywhere-defined, this means
    \be
    \eta({\cal D}(H))={\cal D}(H^\dagger),
    \label{domain1}
    \ee
where ${\cal D}(L)$ denotes the domain of a linear operator
$L:{\cal H}\to{\cal H}$, and $L({\cal S})$ stands for the image of
a subset ${\cal S}\subseteq{\cal H}$ under $L$.

Definition~1 is a direct generalization of the notion of a
self-adjoint operator, for the latter corresponds to a
pseudo-Hermitian operator admitting the identity operator $I:{\cal
H}\to{\cal H}$ as a pseudo-metric.\footnote{We may similarly
generalize the notion of a \emph{symmetric operator}
\cite{reed-simon} by replacing (\ref{domain1}) with $\eta({\cal
D}(H))\subseteq{\cal D}(H^\dagger)$ and requiring that (\ref{ph})
holds in $\eta({\cal D}(H))$.}

In \cite{solombrino}, Solombrino has slightly weakened the
defining condition of a pseudo-Hermitian operator by relaxing the
requirement of the Hermiticity of $\eta$. This leads to the
following notion of weak pseudo-Hermiticity.
\begin{itemize}
\item[]{\bf Definition~2:} A linear operator $H:{\cal H}\to{\cal
H}$ acting in a separable Hilbert space ${\cal H}$ is said to be
\emph{weakly pseudo-Hermitian} if there exists an
everywhere-defined, bounded, invertible, linear map (i.e., a
bounded automorphism) $\eta_w:{\cal H}\to{\cal H}$ satisfying
    \be
    H^\dagger=\eta_w\,H\eta_w^{-1}.
    \label{w-ph}
    \ee
\end{itemize}
Again (\ref{w-ph}) implies
    \be
    \eta_w({\cal D}(H))={\cal D}(H^\dagger).
    \label{w-domain1}
    \ee

The basic motivation for introducing weak pseudo-Hermiticity is
that the Hermiticity of $\eta$ in (\ref{ph}) does not play any
significant role in establishing the spectral characterization
theorem(s) for diagonalizable pseudo-Hermitian operators with a
discrete spectrum \cite{p1,p23,solombrino}. This suggests, at
least for diagonalizable operators with a discrete spectrum, that
pseudo-Hermiticity and weak pseudo-Hermiticity are equivalent
conditions \cite{solombrino}. In \cite{bagchi-quesne}, Bagchi and
Quesne explore the relationship between these two concepts and use
the term ``complementary'' to describe it. Though it is not made
explicit in their analysis, their approach can be consistently
applied only to a restricted class of bounded automorphisms
$\eta$, namely to those for which $\eta+\eta^\dagger$ is also an
automorphism. More recently, Znojil \cite{znojil-pla-2006} has
suggested that considering weak-pseudo-Hermitian Hamiltonians may
provide further insight in the current search for potential
applications of non-Hermitian Hamiltonians in quantum mechanics.

The purpose of this paper is to conduct a careful reexamination of
the relationship between weak pseudo-Hermiticity and
pseudo-Hermiticity for a general not necessarily diagonalizable
linear operator. We will establish the equivalence of these
concepts for a large class of linear operators including all
linear operators that act in a finite-dimensional Hilbert space,
i.e., matrix Hamiltonians.

Before starting our analysis we introduce our conventions and
notation.
\begin{itemize}
\item ${\cal H}$ denotes a separable Hilbert space; \item For any
linear operator $H:{\cal H}\to{\cal H}$, $\fU_H$ stands for the
set of all bounded automorphisms $\eta_w:{\cal H}\to{\cal H}$
satisfying (\ref{w-ph}). Therefore, $H$ is weakly pseudo-Hermitian
if $\fU_H\neq\emptyset$. It is pseudo-Hermitian if $\fU_H$
contains a Hermitian element; \item For any bounded operator
$B:{\cal H}\to{\cal H}$, $\parallel B\parallel$ denotes the norm
of $B$.
\end{itemize}

\section{A Careful Look at Weak Pseudo-Hermiticity}

First we present some useful facts.
\begin{itemize}
\item[]{\bf Proposition~1:} Let $H:{\cal H}\to{\cal H}$ be a
weakly pseudo-Hermitian linear operator. Then for all
$\eta_w,\eta_w'\in\fU_H$, $\eta_w^{-1}\eta_w'$ is a bounded
automorphism commuting with $H$.

\item[]{\bf Proof:} Let $\eta_w,\eta_w'\in\fU_H$. Because both
$\eta_w$ and $\eta_w'$ are bounded, one-to-one, and onto, so are
$\eta_w^{-1}$ and $\eta_w^{-1}\eta_w'$. Furthermore, as shown in
\cite{p1}, in view of (\ref{w-ph}) and its analog satisfied by
$\eta_w'$, we have $[\eta_w^{-1}\eta_w',H]=0$.~~~$\square$

\item[]{\bf Proposition~2:} Let $H:{\cal H}\to{\cal H}$ be a
closed weakly pseudo-Hermitian linear operator and
$\eta_w\in\fU_H$. Then $\eta_w^\dagger\in\fU_H$ provided that
$\eta_w^\dagger{\cal D}(H)={\cal D}(H^\dagger)$. In this case
    \be
    A:=\eta_w^{-1}\eta_w^\dagger
    \label{A=}
    \ee
is a bounded automorphisms commuting with $H$.\footnote{Although
Definitions~1 and 2 do not require $H$ to be a closed operator.
This requirement is necessary to derive many of the useful
properties of pseudo-Hermitian and weakly pseudo-symmetric
operators. Here we need it to assure that $H^{\dagger\dagger}=H$,
\cite{yosida}.}

\item[]{\bf Proof:} Let $\eta_w\in\fU_H$ be such that
$\eta_w^\dagger{\cal D}(H)={\cal D}(H^\dagger)$. $\eta_w$
satisfies (\ref{w-ph}) or equivalently
    \be
    \eta_w H = H^\dagger \eta_w.
    \label{eq1}
    \ee
This in particular implies
    \be
    \eta_w^{-1}{\cal D}(H^\dagger)={\cal D}(H^\dagger\eta_w)=
    {\cal D}(\eta_w H)={\cal D}(H).
    \label{domain-z}
    \ee
Now, take the adjoint of both sides of this
equation.\footnote{Note that for a pair of (densely defined)
linear operators $A,B:{\cal H}\to{\cal H}$ that are not bounded
and everywhere-defined, the relation $(AB)^\dagger=B^\dagger
A^\dagger$ does not hold in general, \cite[\S 7.7]{schechter}.}
Because ${\cal D}(\eta_w H)={\cal D}(H)$ is dense and $\eta_w$ is
bounded and everywhere-defined, we have \cite[\S 7.7]{schechter}
    \be
    (\eta_w H)^\dagger=H^\dagger\eta_w^\dagger,
    \label{eq1-1}
    \ee
or alternatively
    \be
    H^\dagger=(\eta_w H)^\dagger{\eta_w^\dagger}^{-1}.
    \label{eq-1-2}
    \ee
Furthermore, as explained in \cite[\S 7.7]{schechter}, because
${\cal D}(H^\dagger \eta_w)=\eta_w^{-1}{\cal D}(H^\dagger)={\cal
D}(H)$ is dense, ${\cal D}(\eta_w^\dagger H)\subseteq {\cal
D}((H^\dagger \eta_w)^\dagger)$ and
    \be
    (H^\dagger \eta_w)^\dagger\psi=\eta_w^\dagger H\psi~~~~~~~
    {\rm for~all}~~~\psi\in{\cal D}(\eta_w^\dagger H)={\cal
    D}(H).
    \label{domain-z2}
    \ee
This in turn means that
    \be
    (H^\dagger \eta_w)^\dagger{\eta_w^\dagger}^{-1}
    \phi=\eta_w^\dagger H{\eta_w^\dagger}^{-1}\phi~~~~~~~
    {\rm for~all}~~~\phi\in {\eta_w^\dagger}{\cal D}(H).
    \label{domain-z3}
    \ee
Therefore, in view of the hypothesis: ${\eta_w^\dagger}{\cal
D}(H)={\cal D}(H^\dagger)$ and Eqs.~(\ref{eq1}), (\ref{eq-1-2}),
(\ref{domain-z2}), (as envisaged in \cite{bagchi-quesne})
    \be
    H^\dagger=\eta_w^\dagger H{\eta_w^\dagger}^{-1}.
    \label{w-ph-dagger}
    \ee
This together with the fact that the adjoint ($\eta_w^\dagger$) of
a bounded automorphism ($\eta_w$) is a bounded automorphism
establish $\eta_w^\dagger\in{\fU}_H$. The fact that $[A,H]=0$
follows from Proposition~1.~~~$\square$

\item[]{\bf Proposition~3:} Let $H$ and $A$ be as in Proposition~2
and $r_A:=\lim_{n\to\infty}\parallel A^n\parallel^{1/n}$ be the
spectral radius \cite{kato} of $A$. Then the spectrum $\sigma_A$
of $A$ lies in the annulus centered at $0\in\C$ and having as its
inner and outer radii $r_A^{-1}$ and $r_A$, respectively, i.e.,
    \be
    \sigma_A\subseteq
    \left\{z\in\C~|~r_A^{-1}\leq|z|\leq r_A\right\}.
    \label{sigma-range}
    \ee
In particular, $\parallel A\parallel \geq r_A\geq 1$.

\item[]{\bf Proof:} According to Proposition~2, $A$ is a bounded
invertible linear map. This implies that $A^\dagger$ and $A^{-1}$
are bounded operators, and the following identities are satisfied
\cite{kato,schechter}.
    \be
    \sigma_{A^\dagger}=\{\lambda\in\C~|~\lambda^*\in\sigma_A\},~~~~~~
    \sigma_{A^{-1}}=\{\lambda\in\C~|~\lambda^{-1}\in\sigma_A\}.
    \label{sigma}
    \ee
Furthermore, because $A^{-1}={\eta_w^{\dagger}}^{-1}\eta_w$,
$A^\dagger=\eta_w{\eta_w^{-1}}^{\dagger}=\eta_w
A^{-1}\eta_w^{-1}$, and $\eta_w$ is invertible, we have
$\sigma_{A^\dagger}=\sigma_{A^{-1}}$, \cite{halmos}. Combining
this result with (\ref{sigma}), we find that for all
$\lambda\in\sigma_A$, $1/\lambda^*\in\sigma_A$. Next, we recall
that for all $\mu\in\sigma_A$, $|\mu|\leq r_A$. Applying this
inequality for $\mu=\lambda$ and $1/\lambda^*$, we then find
$r_A^{-1}\leq|\lambda|\leq r_A$ for all
$\lambda\in\sigma_A$.\footnote{Because there is always
$\Lambda\in\sigma_A$ such that $|\Lambda|=r_A$ and
$1/\Lambda^*\in\sigma_A$, $\sigma_A$ intersects both the circles
$|z|=r_A$ and $|z|=r_A^{-1}$.} This
establishes~(\ref{sigma-range}). Finally, because the spectrum of
every bounded operator is nonempty, we must have $r_A^{-1}\leq
r_A$, which in turn implies $r_A\geq 1$. The fact that
$r_A\leq\parallel A\parallel$ is well-known,
\cite{kato}.~~~$\square$
\end{itemize}

The following is our main result. It links the equivalence of weak
pseudo-Hermiticity and pseudo-Hermiticity of a large class of
linear operators $H$ with the existence of an $\eta_w\in\fU_H$
such that the unit circle
$S^1:=\{e^{i\varphi}\in\C~|~\varphi\in[0,2\pi)\}$ is not a subset
of $\sigma_A$. Note that Proposition~3 does not rule out this
possibility.

\begin{itemize}
\item[]{\bf Theorem~1:} Let $H:{\cal H}\to{\cal H}$ be a closed
weakly pseudo-Hermitian linear operator acting in a separable
Hilbert space ${\cal H}$. Then $H$ is pseudo-Hermitian, if there
is $\eta_w\in\fU_H$ such that $\eta_w^\dagger{\cal D}(H)={\cal
D}(H^\dagger)$ and the unit circle $S^1$ is not a subset of the
spectrum $\sigma_A$ of $A:=\eta_w^{-1}\eta_w^\dagger$.

\item[]{\bf Proof:} Let $\eta_w\in\fU_H$ be such that
$\eta_w^\dagger{\cal D}(H)={\cal D}(H^\dagger)$, and
$\vartheta\in[0,2\pi)$ be arbitrary. Then according to
Proposition~2, $\eta_w^\dagger\in\fU_H$, and both (\ref{w-ph}) and
(\ref{w-ph-dagger}) hold. Expressing these equations in the form
    \bea
    \eta_w H &=& H^\dagger \eta_w,
    \label{eq1-new}\\
    \eta_w^\dagger H &=& H^\dagger \eta_w^\dagger,
    \label{eq2}
    \eea
multiplying both sides of (\ref{eq1-new}) and (\ref{eq2})
respectively by $ie^{i\vartheta}$ and $-ie^{-i\vartheta}$, and
adding the resulting equations side by side, we find
    \be
    \eta(\vartheta) H=H^\dagger \eta(\vartheta),
    \label{ph-prime}
    \ee
where
    \be
    \eta(\vartheta):=i(
    e^{i\vartheta}\eta_w-e^{-i\vartheta}
    \eta_w^\dagger)~~~~{\rm for~all}
    ~~~~\vartheta\in[0,2\pi).
    \label{eta-prime}
    \ee
The operator $\eta(\vartheta)$ is manifestly Hermitian. It is also
everywhere-defined and bounded, because both $\eta_w$ and
$\eta_w^\dagger$ share these properties. But it needs not be
invertible. We can express $\eta(\vartheta)$ in the form
    \be
    \eta(\vartheta)=-ie^{-i\vartheta}\eta_w(A-e^{2i\vartheta}I),
    \label{eta-prime-2}
    \ee
where $I$ stands for the identity operator acting on ${\cal H}$.
Clearly because $\eta_w$ is invertible, $\eta(\vartheta)$ is
invertible if and only if so is $A-e^{2i\vartheta}I$. By the
definition of the spectrum of a linear operator \cite{reed-simon}
-- \cite{kato}, the latter condition is equivalent to
$e^{2i\vartheta}\notin\sigma_A$. If $S^1\not\subseteq\sigma_A$,
there is $\vartheta_\star\in[0,\pi)$ such that
$e^{2i\vartheta_\star}\notin\sigma_A$. Therefore,
$\eta_\star:=\eta(\vartheta_\star)$ is invertible; $\fU_H$
includes a Hermitian operator $\eta_\star$; and $H$ is
pseudo-Hermitian.~~~~$\square$

\item[]{\bf Corollary:} A linear operator acting in a
finite-dimensional Hilbert space is weakly pseudo-Hermitian if and
only if it is  pseudo-Hermitian.\footnote{As pointed out by the
referee, this is a known result \cite{horn-johnson}.}

\item[]{\bf Proof:} According to Definitions~1 and 2, every
pseudo-Hermitian operator is weakly-pseudo-Hermitian. The converse
holds for an operator acting in a finite-dimensional Hilbert
space, because in this case all the operators are
everywhere-defined (bounded and hence closed) and $\sigma_A$ of
Theorem~1 is a finite set. Hence, it cannot include $S^1$ as a
subset.~~~~$\square$
\end{itemize}

In summary, a weakly pseudo-Hermitian linear operator may fail to
be pseudo-Hermitian, if it acts in an infinite-dimensional space
and for every $\eta_w\in\fU_H$ either $\eta_w^\dagger{\cal
D}(H)\neq{\cal D}(H^\dagger)$ or $S^1\subseteq\sigma_A$ where
$A:=\eta_w^{-1}\eta_w^\dagger$. The latter condition seems to be
very difficult to satisfy.

\section{Examples}

Consider the following bounded automorphism that is employed in
\cite{znojil-pla-2006}.
    \be
    \eta_w=\left(\begin{array}{ccc}
    0&0&{\cal P}\\
    {\cal P}&0&0\\
    0&{\cal P}&0\end{array}\right),
    \label{znojil}
    \ee
where ${\cal P}$ is the usual parity operator acting in $L^2(\R)$,
the Hilbert space is $L^2(\R)\oplus L^2(\R)\oplus L^2(\R)$, a
three-component representation of the state vectors is used, and
$H=(H_{ij})$ is a $3\times 3$ matrix of differential operators
$H_{ij}$ such that ${\cal P}{\cal D}(H)={\cal D}(H)={\cal
D}(H^\dagger)$. It is not difficult to see that
$\vartheta_\star=\frac{3\pi}{2}$ fulfills the conditions of
Theorem~1, and
    \be
    \eta_\star:=\eta(\mbox{$\frac{3\pi}{2}$})=\eta_w+\eta_w^\dagger=
    \left(\begin{array}{ccc}
    0&{\cal P}&{\cal P}\\
    {\cal P}&0&{\cal P}\\
    {\cal P}&{\cal P}&0\end{array}\right)
    \label{eta-znojil}
    \ee
is a genuine pseudo-metric belonging to $\fU_H$. Indeed, it is not
only everywhere-defined, bounded, Hermitian, and one-to-one, but
it is also onto and its inverse is bounded. This can be directly
checked. Alternatively, we may apply Theorem~1 and show that
$S^1\not\subseteq\sigma_A$. It is very easy to compute the
symmetry generator (\ref{A=}):
    \be
    A=\left(\begin{array}{ccc}
    0&0&1\\
    1&0&0\\
    0&1&0\end{array}\right),
    \label{A-znojil}
    \ee
where $1$ is viewed as the identity operator acting in $L^2(\R)$.
Clearly, $\sigma_A=\{1,e^{2i\pi/3},e^{4i\pi/3}\}$. Hence
$S^1\not\subseteq\sigma_A$ and $\eta_\star$ is invertible.

This calculation shows that the systems considered in
\cite{znojil-pla-2006} can be identified with
$\eta_\star$-pseudo-Hermitian Hamiltonians acting in
$L^2(\R)\oplus L^2(\R)\oplus L^2(\R)$ and commuting with $A$,
where $\eta_\star$ and $A$ are respectively given by
(\ref{eta-znojil}) and (\ref{A-znojil}). \emph{These systems can
be studied without any reference to weak pseudo-Hermiticity}.

Another probably more interesting example is $\eta_w:\C^2\to\C^2$
that is defined by its standard matrix representation according to
    \be
    \eta_w=\left(\begin{array}{cc}
    1&1\\
    -1&i\end{array}\right).
    \label{eta-2d}
    \ee
The symmetry generator (\ref{A=}) and the most general Hamiltonian
$H:\C\to\C$ satisfying (\ref{eq1}) have the following standard
matrix representations
    \bea
    A&=&\left(\begin{array}{cc}
    i&0\\
    1-i&-1\end{array}\right)=-iM_1-M_2,
    \label{A-2d}\\
    H&=&\left(\begin{array}{cc}
    a&0\\
    ib&a+b\end{array}\right)=aI+b M_1,
    \label{2d}
    \eea
where
    \[M_1:=\left(\begin{array}{cc}
    0 & 0 \\
    i& 1\end{array}\right),~~~~~~
    M_2:=\left(\begin{array}{cc}
    1 & 0 \\
    -i& 0\end{array}\right),\]
$I$ is the identity matrix, and $a,b\in\R$ are arbitrary. Clearly,
$A$ and $H$ commute for all $a,b\in\R$.

We can also easily compute $A-e^{2i\vartheta}I$. It turns out to
be non-invertible only for
$\vartheta=\frac{\pi}{4},\frac{\pi}{2},\frac{5\pi}{4},\frac{3\pi}{2}$.
This in turn means that $\eta(\vartheta)$ is non-invertible for
these four values of $\vartheta$. In particular,
$\eta(\frac{3\pi}{2})=\eta_w+\eta_w^\dagger$ that is considered in
\cite{bagchi-quesne} is not invertible.\footnote{The possibility
that given an invertible operator $\eta_w$ the operators
$\eta_w\pm\eta_w^\dagger$ may fail to be invertible seems to be
overlooked in \cite{bagchi-quesne}.}

In general, $\eta(\vartheta)$ has the following explicit form
    \be
    \eta(\vartheta)=2\,c\left(\begin{array}{cc}
    -t & i\\
    -i & -1\end{array}\right),
    \label{eta-alpha=}
    \ee
where $c:=\cos\vartheta$ and $t:=\tan\vartheta$. In terms of $c$
and $t$ the invertibility condition:
$\vartheta\notin\{\frac{\pi}{4},\frac{\pi}{2},\frac{5\pi}{4},
\frac{3\pi}{2}\}$, takes the simple form: $c\neq 0$ and $t\neq 1$.

Having obtained an infinite class of pseudo-metric operators
$\eta(\vartheta)$ that render the Hamiltonians of the form
(\ref{2d}) pseudo-Hermitian, we can construct the following family
of symmetry generators \cite{p1,jmp-2003}
    \be
    A(r,t_1,t_2):=\eta(\vartheta_2)^{-1}\eta(\vartheta_1)
    =r\left(\begin{array}{cc}
    1-t_1 & 0 \\
    i(t_1-t_2)&1-t_2
    \end{array}\right)=r(I-t_2 M_1-t_1 M_2),
    \label{A12}
    \ee
where
$r:=\frac{\cos\vartheta_1}{\cos\vartheta_2-\sin\vartheta_2}\neq 0$
and $t_i:=\tan\vartheta_i\neq 1$ for $i=1,2$. Comparing (\ref{2d})
and (\ref{A12}), we see that the only nontrivial symmetry
generator for the system is $M_2$. We could reach the same
conclusion using (\ref{A-2d}).

Finally, we note that $\eta(\alpha)$ is positive-definite whenever
$c<0$ and $t>1$ which corresponds to
$\frac{5\pi}{4}<\vartheta<\frac{3\pi}{2}$. In particular, $H$ is
pseudo-Hermitian with respect to a set of positive-definite metric
operators. According to \cite{p23}, this implies that it is
quasi-Hermitian \cite{quasi} and has real eigenvalues. The latter
is easily seen from (\ref{2d}) where the eigenvalues appear as
diagonal entries.

\section{Concluding Remarks}

In this paper, we have examined the relation between the notions
of pseudo-Hermiticity and weak pseudo-Hermiticity. We have found a
sufficient spectral condition that ensures whether a given weakly
pseudo-Hermitian operator is pseudo-Hermitian. This condition
which is not sensitive to the diagonalizability of the operator in
question is trivially satisfied in finite-dimensional Hilbert
spaces. Hence weak pseudo-Hermiticity and pseudo-Hermiticity are
equivalent in finite dimensions. This equivalence extends to a
large class of operators acting in infinite-dimensional Hilbert
spaces. Our general results seem to indicate that further
investigation of weak pseudo-Hermiticity is not likely to produce
any substantial insight in the current study of the possible
applications of non-Hermitian Hamiltonians in quantum mechanics.

\subsection*{Acknowledgment}

During the course of this work I have benefitted from helpful
discussions with Varga Kalantarov. I would also like to thank the
anonymous referee for bringing Ref.~\cite{horn-johnson} to my
attention and for correcting an error in a previous version of the
paper.


\ed